\documentclass[twocolumn]{webofc}
\usepackage[varg]{txfonts}   % Web of Conferences font
\usepackage{amsmath}
\usepackage{graphicx}
\usepackage{float}
\usepackage[colorlinks = true,linkcolor = blue,urlcolor  = blue,citecolor = blue,anchorcolor = blue]{hyperref}
\usepackage{natbib}
\usepackage{microtype}
\begin{document}
\title{Twin stars: probe of phase transition from hadronic to quark matter}

\author{\firstname{Themistoklis} \lastname{Deloudis}\inst{1}\fnsep\thanks{\email{tdeloudi@physics.auth.gr}} \and
        \firstname{Polychronis} \lastname{Koliogiannis}\inst{1}\fnsep\thanks{\email{pkoliogi@physics.auth.gr}} \and
        \firstname{Charalampos} \lastname{Moustakidis}\inst{1}\fnsep\thanks{\email{moustaki@auth.gr}}
        % etc.
}

\institute{Department of Theoretical Physics, Aristotle University of Thessaloniki, 54124 Thessaloniki, Greece}

\abstract{%
  In agreement with the gravitational-wave events which are constantly increasing, new aspects of the internal structure of compact stars have come to light. A scenario in which a first order transition takes place inside these stars is of particular interest as it can lead, under conditions, to a third gravitationally stable branch (besides white dwarfs and neutron stars). This is known as the twin star scenario. The new branch yields stars with the same mass as normal compact stars but quite different radii. In the current work, we focus on hybrid stars undergone a hadron to quark phase transition near their core and how this new stable configuration arises. Emphasis is to be given especially in the aspects of the phase transition and its parametrization in two different ways, namely with Maxwell construction and with Gibbs construction. Qualitative findings of mass-radius relations of these stars will also be presented.
}
\maketitle
\section{Introduction}
Compact stars yield the most prominent natural laboratories for the study of exotic forms of matter~\cite{Glendenning-1997,Haensel-2007}. Recently discovered pulsars alongside with gravitational waves detection, such as GW170817, have revealed new aspects of the internal structure of these stars, mainly in terms of their composition~\cite{Christian-2018,Montana-2019}. Whilst the equation of state of nuclear matter is well established up to nuclear saturation density, one encounters the challenge of describing matter in fairly higher densities realized in the interior of these stars. At these densities the type of matter is yet to be determined and in turn, the construction of stellar models to agree with the aforementioned observations is still an open issue. Possible candidates are strange quark stars composed of deconfined quarks, pure neutron stars composed by hadrons, and hybrid stars composed by hadronic outer shells and cores of deconfined quarks. In the present article the latter case is claimed.

The idea of a third family of compact stars and in particular, the connection with the possibility to be a signature of a strong phase transition in the interior of the star, was introduced first by  Gerlach~\cite{Gerlach-1968}. Later on, K{\"a}mpfer worked also on this issue~\cite{Kampfer-1981a,Kamfer-1981b}. Glendenning and Kettner introduced the term ``twins" in their paper~\cite{Glendenning-2000}, while at the same time Schertler et al.~\cite{Schertler-2000} worked out that idea in detail. However, in all the previous studies, the maximum mass was approximately at the canonical binary pulsar mass $1.4~M_{\odot}$. The revival of the idea of the twin stars started a few years later by Blaschke et al.~\cite{Blaschke-2013,Castillo-2013}. Specifically, in the mentioned papers is suggested that high-mass twin stars, once detected by simultaneous mass and radius measurements, could provide the evidence for a strong first-order phase transition in cold matter, which then would imply the existence of at least one critical endpoint in the quantum chromodynamics phase diagram. Moreover, in the same work is also presented, for the first time, examples of equations of state that would not only provide twin solutions, but also fulfill the constraint on the maximum mass from the existence of pulsars as heavy as $2~M_{\odot}$~\cite{Blaschke-2013,Castillo-2013}. The previous idea was elaborated by Benic et al.~\cite{Benic-2015} (see also Ref.~\cite{Castillo-2015}). A systematic Bayesian analysis of the new twin star equation of state with observational constraints was presented in Ref.~\cite{Castillo-2016}. Finally, the analysis of the robustness of twin solutions against changing the Maxwell to a mixed phase construction, and the formation of structures in the mixed phase due to the interplay of the surface tension and Coulomb interaction effects, have been also considered respectively in Refs~\cite{Ayriyan-2018,Maslov-2019}.   

Stars of this branch are expected to have masses in the same range as normal neutron stars, yet fairly smaller radii. The existence of such stars is a strong indication that a HQPT is a physical reality, a result of utmost importance especially in the study of dense matter physics.
A  study of the HQPT is presented here, examining the conditions under which the twin star configuration arises. The compatibility with the mass and radius constraints as they are formed through up to date observations is also considered.

%The article is organized as follows. In section~\ref{sec:2} hybrid star configurations which undergone a sharp or a non-sharp %HQPT, are discussed in detail. In section~\ref{sec:3}, results of our work on non-rotating hybrid stars with particular %emphasis on the occurrence of twin stars is presented while section~\ref{sec:4} concerns our conclusions.

%%%%%%%%%%%%%%%%%%%%%%%%%%%%%%%%
\section{Theoretical Framework}\label{sec:2}
%%%%%%%%%%%%%%%%%%%%%%%%%%%%%%%%%%%
Examining the configuration of a compact star is efficiently achieved by separating the interior into regions. Generally, it is suitable to claim a hadronic phase composing the outer shells, a core of deconfined quarks and a connecting region embodying the phase transition. As it will become clear  in the following, this last region can be omitted yielding a sharp phase transition or, as it is commonly called, a Maxwell construction (MC). Oppositely, Gibbs construction (GC) assumes the existence of this transitional region, which is commonly referred as the mixed phase. In each case, the phase transition is assumed to be of first order~\cite{Christian-2018,Montana-2019,Glendenning-2000,Alvarez-2017,Bejger-2017,Bhattacharyya-2010,Endo-2006,Christian-2019,Christian-2021,Alford-2013,Han-2019}. 

%%%%%%%%%%%%%%%%%%%%%%%%%%%%%%%%%%%%%
\subsection{Hadronic phase}
%%%%%%%%%%%%%%%%%%%%%%%%%%%%%%%%%%
In the outer layers of a compact star, matter is expected to be mainly hadronic. It can either consist solely of nucleons or other hadrons, usually hyperons. In our study the APR model introduced in Ref.~\cite{Akmal-1998} combined with the momentum dependent interactions (MDI) parametrization in the way presented in Refs.~\cite{Koliogiannis-2020,Koliogiannis-2021} is applied. Hence, no hadrons other than protons and neutrons are assumed. The selection of this particular hadronic model is due to the numerous advantages provided, namely the reproduction of symmetric nuclear matter properties, value and slope of symmetry energy at the saturation density and the agreement with limits provided by experimental predictions. Furthermore, state-of-the-art calculations considering high densities, as well as calculations of the Chiral model considering pure neutron mater are successfully reproduced. Lastly, as far as the results are concerned, this model can reproduce maximum masses of neutron stars even higher than those observed~\cite{Koliogiannis-2020,Koliogiannis-2021,Arzoumanian,Antoniadis,Cromartie}.

%%%%%%%%%%%%%%%%%%%%%%%%%%
\subsection{Quark phase}
%%%%%%%%%%%%%%%%%%%%%%%%%%%%%%%%
Deconfined quark matter composing the stellar core is in general difficult to formulate~\cite{Baym-2018}. One needs sophisticated models to study it in detail. Nevertheless, one possibility is to mimic these models by assuming constant value of the speed of sound throughout the quark phase. This is a robust feature on top of being practical, as it simplifies the parametrization of the energy density expression. Apparently, the speed of sound $c_{\rm s}^2=\partial p/\partial \mathcal{E}$ is subjected to the constraint $c_{\rm s}\leq1$ in all possible cases. Its exact value plays a decisive role on the final configuration that the star will achieve.

%%%%%%%%%%%%%%%%%%%%%%%%%%%%%%%%
\subsection{Phase transition}
%%%%%%%%%%%%%%%%%%%%%%%%%%%%%%%%%%%
As the density of stellar matter soars above the nuclear saturation density, a phase transition must take place in order for the quark core to occur. A common approach to describe the former is with a polytropic equation of state of the form~\cite{Christian-2018}
\begin{equation}
	p=K\rho^\Gamma,
	\label{polyt-1}
\end{equation}
where $p$ stands for pressure, $\rho$ for baryon density and $K,\Gamma$ are constants. Through Eq.~\eqref{polyt-1} it is easy to distinguish between a sharp transition (MC), by imposing $\Gamma=0$, and a non-sharp transition (GC), assuming $\Gamma\neq0$.

A phase transition under constant pressure, as in MC, leads to a specific structure of the star. Due to the monotonic reduction of the pressure relative to the radius, points of equal pressure must be mapped onto the same radial coordinate inside the star~\cite{Glendenning-1997}. That is, the hadronic phase of the outer shells and the quark phase of the core are essentially in direct contact with each other at the point of transition, as a result of the absence of a finite transitional region. This is not the case in GC where the configuration includes this region and a regular behaviour of the pressure inside the star. In the following, these two cases are examined in detail.

%%%%%%%%%%%%%%%%%%%%%%%%%%%%%%%%%%%%%
\subsubsection{Maxwell construction}
%%%%%%%%%%%%%%%%%%%%%%%%%%%%%%%%%%%%%%%%%
An abrupt phase change yields discontinuity in at least one physical quantity. In this study the energy density is of primary concern, claiming the form~\cite{Christian-2018,Montana-2019}
\begin{eqnarray}
	\hspace{-0.9cm}
	\mathcal{E}(p)&=&\left\{
		\begin{array}{ll}
			\mathcal{E}_{\rm APR+MDI}(p), \quad  p<p_{\rm tr} & \\
			\\
			\mathcal{E}(p_{\rm tr})+\Delta \mathcal{E} + (p-p_{\rm tr})~c_{\rm s}^{-2}, \quad  p\geq p_{\rm tr} & \
		\end{array}
	\right.
	\label{MC-1}
\end{eqnarray}
where $\mathcal{E}(p)$ denotes the energy density, $p$ the pressure, $c_{\rm s}$ the speed of sound, and $\Delta \mathcal{E}$ is the magnitude of the energy density jump at the transition point. Subscript $``{\rm tr}"$ denotes the corresponding quantity at this point. It should be clear that the first line in Eq.~\eqref{MC-1} refers to the hadronic phase while the second one refers to the quark phase. 

By definition, MC is characterized by a precondition which, when satisfied, leads to an unstable regime essential for the third branch of stable hybrid stars to arise~\cite{Christian-2018,Montana-2019,Alvarez-2017}. This constraint, first shown by Seidov~\cite{Seidov-1971}, is expressed as
\begin{equation}
	\Delta \mathcal{E} \geq \frac{3p_{\rm tr}+\mathcal{E}(p_{\rm tr})}{2}.
	\label{Seid-1}
\end{equation}
The star will immediately become unstable after its central pressure reaches the value $p_{\rm tr}$. Otherwise, a stable state occurs. Therefore, when seeking twin stars with MC one should always ensure that Eq.~\eqref{Seid-1} is satisfied.

%%%%%%%%%%%%%%%%%%%%%%%%%%%%%%%%%%%
\subsubsection{Gibbs construction}
%%%%%%%%%%%%%%%%%%%%%%%%%%%%%%%%%%%%%%%%
In a case of a non-sharp HQPT (GC), a finite region to embody the transition is implied. The mixed phase of this transition is composed, as its name suggests, by intermittent domains of pure hadronic and quark phases~\cite{Bhattacharyya-2010,Endo-2006}. Contrary to MC, no discontinuities in energy density appear, giving rise to the profile
\begin{eqnarray}
	\hspace{-0.9cm}
	\mathcal{E}(p)&=&\left\{
		\begin{array}{ll}
			\mathcal{E}_{\rm APR+MDI}(p), \quad p\leq p_{\rm tr} & \\
			\\
			\Lambda p^{1/\Gamma}+p~(\Gamma-1)^{-1}, \quad p_{\rm tr}\leq p\leq p_{\rm css} & \\
			\\
			\mathcal{E}(p_{\rm css})+(p-p_{\rm css})~c_{\rm s}^{-2}, \quad p\geq p_{\rm css}. & \
		\end{array}
	\right.
	\label{GC-1}
\end{eqnarray}
Again, the energy density is denoted by $\mathcal{E}(p)$, the pressure by $p$, the speed of sound by $c_{\rm s}$, while $\Lambda$ and the polytropic index $\Gamma$ are constants. Subscript $``{\rm \rm css}"$ denotes the corresponding quantity at the start of the quark phase. In addition, the continuity at the two transition points imposes
\begin{equation}
	\Lambda=\frac{\mathcal{E}(p_{\rm tr})-p_{\rm tr}~(\Gamma-1)^{-1}}{p_{\rm tr}^{1/\Gamma}},
\end{equation}
and
\begin{equation}
	\mathcal{E}(p_{\rm \rm css})=\left(\frac{p_{\rm \rm css}}{p_{\rm tr}}\right)^{1/\Gamma}\left(\mathcal{E}(p_{\rm tr})-\frac{p_{\rm tr}}{\Gamma-1}\right)+\frac{p_{\rm \rm css}}{\Gamma-1},
\end{equation}
Apparently, the first line in Eq.~\eqref{GC-1} corresponds to the hadronic outer shells, the second one to the mixed phase, embodying the HQPT, and the third one to the quark core.
\begin{figure}
	\includegraphics[width=\columnwidth]{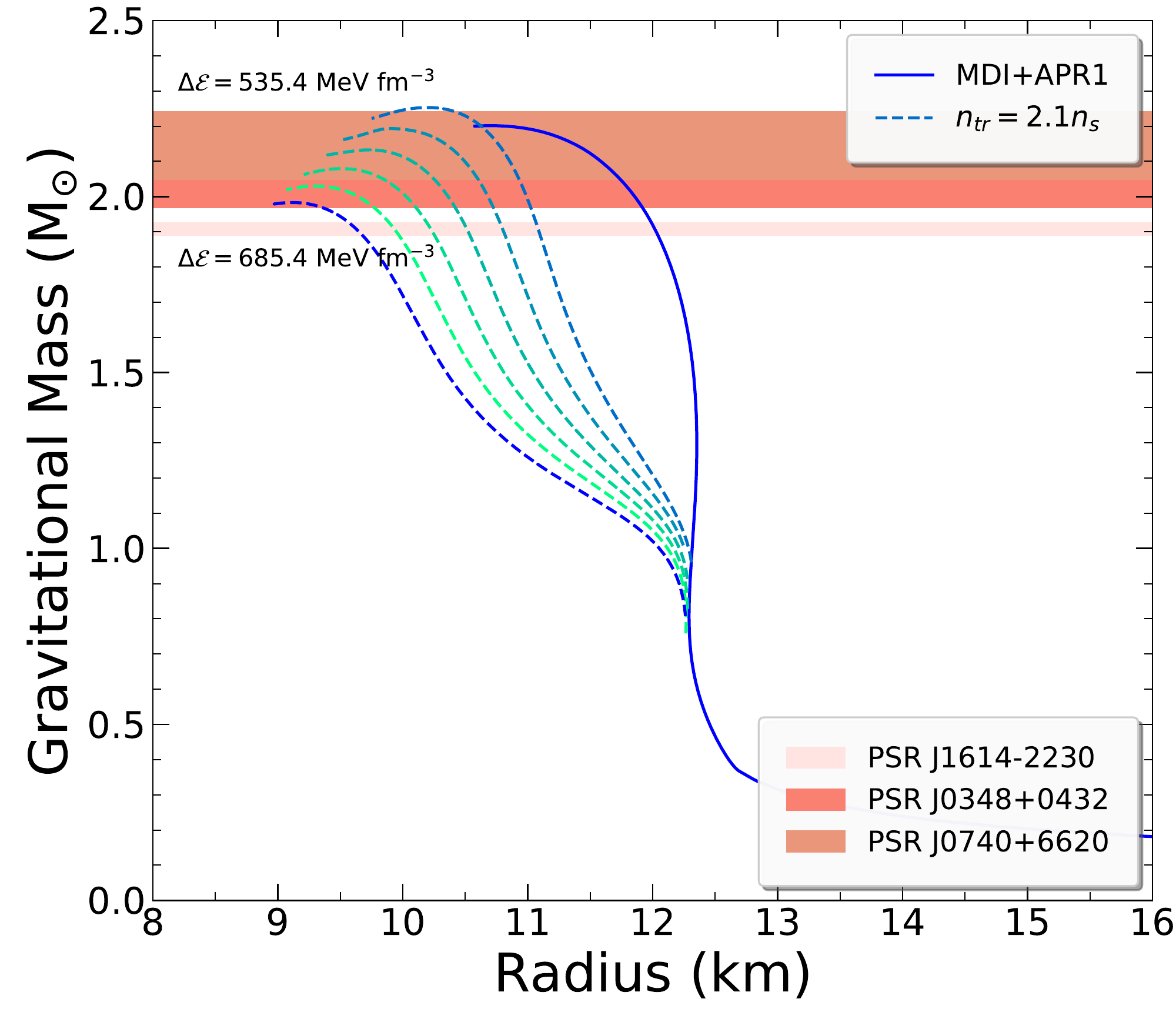}
	\caption{Mass-Radius diagram in the case of a sharp transition at $n_{\rm tr}=2.1~n_{\rm s}$ (dashed lines) for various values of the energy density jump $\Delta \mathcal{E}$. The solid line refers to the absence of a phase transition and is added for comparison. The coloured strips near the top of the figure denotes the range (uncertainty) around the maximum observed mass values of three pulsars~\cite{Arzoumanian,Antoniadis,Cromartie}.}
	\label{fig:1}
\end{figure}
An immediate consequence of GC is that an energy density jump is not defined as in MC. Therefore, in comparing the energy density change due to the transition in the two constructions one needs to use the rise of the energy density throughout the mixed phase in GC as a counterpart of $\Delta \mathcal{E}$ of MC. It is immediately seen that
\begin{eqnarray}
	\Delta \mathcal{E}_{\rm GC}&=&\mathcal{E}(p_{\rm tr})\left[\left(\frac{p_{\rm \rm css}}{p_{\rm tr}}\right)^{1/\Gamma}-1\right]\nonumber \\ &+&\frac{1}{\Gamma-1}\left[p_{\rm \rm css}-\left(\frac{p_{\rm \rm css}}{p_{\rm tr}}\right)^{1/\Gamma}p_{\rm tr}\right].
\end{eqnarray}
Furthermore, in GC no constraint similar to Eq.~\eqref{Seid-1} is present. In order to achieve twin star configuration only the appropriate choice of the free parameters is needed.

%%%%%%%%%%%%%%%%%%%%%%%
\section{Results} \label{sec:3}
%%%%%%%%%%%%%%%%%%%%%%%%%%
Applying our aforementioned assertions on a non-rotating stellar model a first approach of the phenomenon is achieved and the basis is set for its study in rotating models. Remarkably enough, a HQPT introduces only two free parameters in each construction. Specifically, these correspond to the critical pressure $p_{\rm tr}$ of the hadronic phase and the magnitude of the discontinuity $\Delta \mathcal{E}$ in the case of MC, or the critical pressure $p_{\rm tr}$ and the mixed phase interval, commonly described by the pressure difference of its boundaries, $\Delta p$, in the case of GC. Other free parameters may exist overall, though not introduced by the implementation of a HQPT. For example, the value of the energy density at the transition point depends on the hadronic equation of state that is used.  

Figure~\ref{fig:1} is a M-R diagram visualising the case of a sharp transition taking place at $n_{\rm tr}=2.1n_{\rm s}$, where $n$ is the baryon density and the subscripts $``{\rm tr}"$ and $``{\rm s}"$ denote its value at the transition and the saturation point, respectively.
\begin{figure}
	\includegraphics[width=\columnwidth]{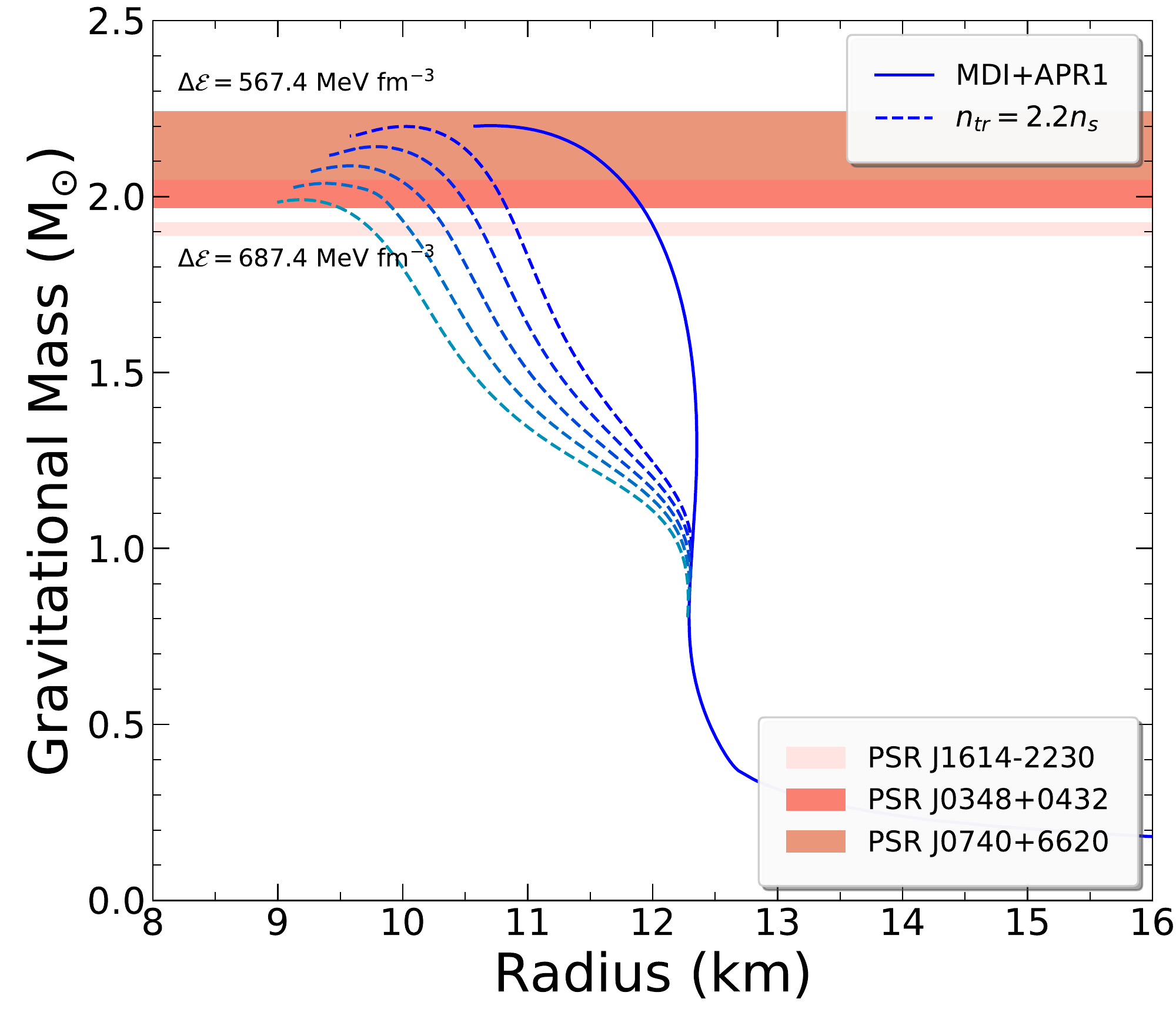}
	\caption{Mass-Radius diagram in the case of a sharp transition at $n_{\rm tr}=2.2~n_{\rm s}$ (dashed lines) for various values of the energy density jump $\Delta \mathcal{E}$. The solid line refers to the absence of a phase transition and is added for comparison. The coloured strips near the top of the figure denotes the range (uncertainty) around the maximum observed mass values of three pulsars~\cite{Arzoumanian,Antoniadis,Cromartie}.}
	\label{fig:2}
\end{figure}
In examining Figure~\ref{fig:1}, the branching due to the existence of a HQPT is immediately evident. The change of $\Delta{\cal E}$ leads to a different position of the second minimum but the shape of the  second branch remains almost  identical. Moreover, it is worth  to notice that higher values of $\Delta{\cal E}$ lead to smaller radii for the corresponding maximum mass configurations. A possible explanation is that larger $\Delta \mathcal{E}$ yield heavier stellar cores with an accompanying greater gravitational pull on the outer shells, as suggested in Ref.~\cite{Montana-2019}. These conclusions are in agreement with Ref.~\cite{Christian-2018}. As predicted by theory, twin star branches yield fairly smaller radii compared with the normal neutron star branch for each value of stellar mass. It is also clear that the higher the energy density jump at the transition point, the lower the resulting stellar mass. Another important remark is that the parametrization used for the HQPT reproduces the maximum masses imposed by observations~\cite{Christian-2018,Montana-2019,Alvarez-2017,Arzoumanian,Antoniadis,Cromartie}. To explicitly show that, the masses of three pulsars belonging to different binary systems are included in Figure~\ref{fig:1}. 

The last point one needs to remark in that figure is the fact that the transition occurs at baryon density equal to 2.1 times the saturation density. We display also the Figure~\ref{fig:2}, in order to study the effect of the location of the transition density on the various twin stars  branches.

We observed that all of the fundamental properties mentioned above are still valid for $n_{\rm tr}=2.2n_{\rm s}$. The only difference is that in this case the point of branching is slightly pushed towards higher masses, a fact that obviously does not affect the agreement with the observations. Lastly, it should be noted that the results for the MC were derived by assuming $c_{\rm s}=1$.

It is worth noting that the case of a non-sharp phase transition (GC) should not be omitted, at least in a form of reference. Gibbs construction yields a more complicated configuration of the compact star. This difficulty has resulted in not many publications reviewing this case in detail. We are currently working on the twin star scenario with GC, with the results yet in progress. We are confident that the extraction of these results will allow not only for a comparison and discrimination between the two constructions, but also to the conclusion on which of the two is the most physically plausible.

%%%%%%%%%%%%%%%%%%%%%%%%%%%%%%%%%%%
\section{Conclusions and outlook} \label{sec:4}
%%%%%%%%%%%%%%%%%%%%%%%%%%%%%%%%%%%%%%%
Analyzing the twin star scenario, it is primarily seen that two ways to formulate the HQPT, MC and GC, are dominant. These seem to cover every plausible scenario for the transition. It has to be noted that configurations achieved with MC are  simpler than GC. However, in MC case  the  condition Eq.~\eqref{Seid-1} must also  be fulfilled. Results posed in Figures~\ref{fig:1} and~\ref{fig:2} with MC show clearly the differentiation of cases in a compact star with and without a phase transition. The formulation achieved for the former case reproduces masses and radius in agreement with observations in a satisfying interval of values of the baryon density at the transition point. For a more complete study, the same results are expected to be deducted for GC.

Future considerations may include:   (a) designating the conditions under which the twin star branch and the normal neutron star branch are separated; (b) achieving plausible configurations with $c_{\rm s}^2=1/3$, a value which is closer to reality than causality; and (c) examining the phenomenon in the frame of rotating stellar models.

\begin{acknowledgement}
The authors would like to thank Prof. D. Blaschke for his useful insight  and comments.
\end{acknowledgement}

\end{document}